# Situated Epistemic Infrastructures:<br>A Diagnostic Framework for Post-Coherence Knowledge

Matthew Kelly

## Epistemic Crisis and Infrastructural Visibility

LLMs, as epistemic infrastructures, expose the fragility of contemporary knowledge systems more starkly than decades of theoretical critique. As ChatGPT and similar systems bypass traditional citation practices, generate seemingly authoritative responses from opaque training data, and compete with human experts for epistemic authority in academic and professional spaces, the organizing assumptions of scholarly communication have become starkly visible. What once appeared as coherent systems of knowledge production now fracture into overlapping, contested, and rapidly shifting networks of algorithmic inference, human interpretation, and institutional response.

This moment does not signal the failure of individual communities or institutions, but a deeper disruption: the destabilization of epistemic coherence as an organizing condition. No single domain governs the trajectory of LLM-mediated knowledge. Instead, knowledge travels through hybrid infrastructures—search engines, chatbots, citation managers, academic databases—each with its own temporal rhythms, training biases, and authority claims. The result is a form of epistemic life shaped less by stable communities than by what Susan Leigh Star (1999) recognized as the active mediation of material–symbolic infrastructures that operate across institutional, technological, and temporal boundaries, now amplified through computational systems that simulate coherence while obscuring their epistemic foundations.

## The Situated Epistemic Infrastructures Framework and Temporal Ethics

The Situated Epistemic Infrastructures (SEI) framework offers a way to analyze how knowledge becomes authoritative through the interaction of infrastructural, institutional, and temporal forces. Rather than assuming stable scholarly communities or coherent domains, SEI focuses on how credibility is mediated across hybrid systems—human and machine, symbolic and material—under conditions of epistemic disruption. The framework proposed here embodies what might be called temporal ethics—recognition that theoretical frameworks achieve their highest expression when they consciously prepare for their own transcendence. Rather than claiming permanent authority over epistemic analysis, SEI creates conditions for more sophisticated future approaches to emerge through conceptual hospitality—theoretical architecture that can accommodate intellectual developments we cannot currently anticipate. This extends Hans-Georg Gadamer's (2004) "fusion of horizons" to encompass technological mediation and temporal dynamics beyond human-centered hermeneutics. This approach operates through stewardship rather than ownership of conceptual development, creating conditions for ideas to develop according to their own logic while accepting responsibility for the quality of initial creative work without claiming permanent authority over subsequent theoretical evolution.

This marks what can be termed *the post-coherence condition*: a state in which the idea of neatly bounded epistemic communities no longer adequately explains how knowledge is produced, validated, or organized. Traditional models—especially in information science—have assumed that knowledge organization can be mapped through coherent domains, defined by stable communities of practice, shared paradigms, and cumulative rationality. But under contemporary conditions of technological acceleration, platform convergence, and institutional reconfiguration, these assumptions encounter the *infrastructural inversion* (Bowker, 1994)—the moment when supposedly neutral technical arrangements reveal themselves as active agents in shaping epistemic possibilities. This process involves what Jesper Simonsen et al. (2020, p. 115) term "the gestalt switch of shifting attention from the activities invisibly supported by an infrastructure to the activities that enable the infrastructure to function and meet desired needs for collaborative support."

SEI adopts a mode of theoretical foresight, aiming not at prediction but at cultivating analytical flexibility for conditions that have yet to emerge. The framework draws inspiration from Barben et al.'s (2008) work on anticipatory governance while rejecting technocratic approaches in favor of what might be called "epistemic anticipation"—preparing conceptual resources rather than predicting specific outcomes. This involves temporal gift-giving—creating conceptual resources available for future intellectual work without specifying exactly how they will be used. In developing diagnostic tools for epistemic breakdown, the framework addresses current crises while also preparing conceptual resources for challenges that have yet to fully emerge. The framework becomes a contribution to future theoretical possibility rather than merely a solution to present analytical problems, participating in what might be called a temporal gift economy where concepts are created for future thinkers encountering challenges we cannot anticipate.

## Beyond Coherence: The Epistemic Condition Today

For decades, knowledge organization research has relied on the notion of epistemic coherence: the idea that knowledge is stabilized through shared vocabularies, disciplinary boundaries, and collectively maintained standards of evidence and authority. This assumption underpins models such as Birger Hjørland's (2002) domain analysis, which focus on how different communities construct and maintain their own coherent knowledge systems through what Bourdieu (1991) would recognize as shared "habitus"—embodied dispositions which are experienced as natural, even though they are socially constructed (Daston & Galison, 2007; Rauch, 2021).

Today's epistemic condition is defined by instability, not coherence. Technological acceleration, platform-driven communication, the proliferation of AI-generated content, and increasing demands for public participation in science have destabilized longstanding modes of knowledge legitimation. The tools of 20th-century knowledge organization—controlled vocabularies, bibliometric hierarchies, peer-reviewed workflows—no longer map cleanly onto the way knowledge is produced or trusted. As epistemic arrangements hybridize, new forms of coordination and conflict emerge through processes that Gabriel Tarde (1903) anticipated as "imitation and suggestion," which create waves of adoption across social networks.

## Ecological Understanding of Theoretical Development

Individual epistemic frameworks participate in larger conceptual ecosystems that evolve according to their own temporal logic. This ecological understanding builds on Crawford S. Holling's (1973) work on ecological resilience while extending it to address conceptual rather than biological systems, suggesting that theoretical frameworks require adaptive capacity rather than static stability. SEI operates through co-evolution with material conditions—AI systems, digital platforms, institutional pressures—rather than pure conceptual development. Theoretical innovation becomes less mechanical and more atmospheric—more like conceptual climate change than linear transmission. The framework emerges precisely through engagement with AI collaboration, digital platforms, and contemporary epistemic crises while potentially transforming them through theoretical intervention.

Contemporary knowledge production operates through cascading effects, where changes in one infrastructural component generate unpredictable transformations across entire epistemic systems. Scholars collaborate on Slack before publishing; open datasets often surpass narrative reviews; preprint citations outpace journals; and expertise is contested publicly before disciplinary ratification. Machine learning models trained on academic literature generate research proposals while social media algorithms determine which scientific findings reach policymakers. These automated processes create what Frank Pasquale (2015) calls "black box" decision-making, where knowledge circulation follows logics that remain opaque to users and reshape the conditions of scholarly authority itself.

## Infrastructural Mediation as Epistemic Foundation

Infrastructures are often understood in technical terms: networks, systems, databases. But following Star's foundational insight, infrastructures are also semiotic and institutional arrangements that organize meaning, distribute authority, and coordinate action. Star (1999, p. 382) observed that "infrastructure becomes visible upon breakdown"—an insight that applies with particular force to knowledge infrastructures. When algorithms misfire, platforms crash, or concepts lose their interpretive clarity through symbolic drift—the process by which legitimating terms become detached from their original practical commitments while retaining their authority—the infrastructural scaffolding of epistemic life suddenly becomes visible.

Building on Star's insight and drawing from actor-network theory's attention to the co-constitution of human and non-human actors (Latour, 2005), SEI treats infrastructure as the active medium through which epistemic practices take shape, rather than as a neutral backdrop. These infrastructures include platforms like arXiv and PubMed, symbolic compressions like "evidence-based," institutional norms like peer review, and what W. Lance Bennett and Alexandra Segerberg (2013) term "power signatures"—here extended to analyze the distinctive modes through which different infrastructural arrangements exercise and maintain epistemic authority. Together, they shape what counts as knowledge, who counts as a knower, and how epistemic legitimacy is distributed through processes that often remain invisible until breakdown occurs.

Rather than focusing on domains or classifications, SEI asks: What infrastructures make beliefs possible? How do these infrastructures maintain stability—or enable transformation—under pressure? What happens when they fail? This reframing shifts the lens from coherence to mediation: from internal alignment to the systems that allow, constrain, and reshape it.

## Situatedness and Reflexivity with Hermeneutic Patience

The SEI framework preserves domain analysis's crucial insight about situatedness while extending it to address infrastructural conditions. Following Donna Haraway's (1988) insistence that all knowledge is "situated" rather than universal, SEI emphasizes that knowledge infrastructures operate differently across contexts, communities, and moments in time. There is no universal infrastructure; there are only situated configurations that must be studied on their own terms, with attention to their local dynamics and global entanglements.

SEI requires a form of hermeneutic patience—dispersed attentiveness that allows peripheral insights to emerge and conceptual resonances to develop slowly rather than forcing predetermined outcomes. This concept extends Paul Ricoeur's (1981) hermeneutic philosophy while incorporating insights from infrastructure studies about the temporal dimensions of sociotechnical arrangements that exceed traditional hermeneutic frameworks. This conceptual hospitality creates space for ideas to develop according to their own logic rather than imposing external demands for premature clarity or immediate application. The approach involves maintaining interpretive availability to whatever emerges from theoretical processes, preserving surplus conceptual capacity—theoretical depth that exceeds immediate analytical requirements while remaining available for future activation.

The framework also emphasizes reflexivity—the need for the field of knowledge organization to develop diagnostic tools that can detect when its own methods are no longer adequate to the world they aim to describe. Drawing from Annmarie Mol's (2002) work on ontological multiplicity and the need for methodological sensitivity to complexity, SEI is both a theoretical framework and a practical orientation: it provides tools for understanding infrastructures, but also for designing, repairing, and adapting them under conditions of uncertainty and change.

## From Domain Analysis to Epistemic Infrastructures

The reconfiguration of knowledge production during COVID-19 crystallized deeper structural tensions that have long been reshaping epistemic life. Traditional approaches to knowledge organization—especially domain analysis—emerged in a world where disciplinary boundaries were relatively stable, institutional hierarchies durable, and epistemic rhythms were comparatively slower than today, creating conditions that allowed consensus to coalesce through what Randall Collins (2004) describes as "interaction ritual chains": recurring performances that generate emotional solidarity around shared standards. These conditions no longer hold.

### The Situated Breakthrough—and Its Limits

Hjørland's (2002) domain analysis revolutionized knowledge organization by asserting that classification must reflect the practices, values, and perspectives of epistemic communities rather than appeal to universal logical categories. It introduced a richly sociocultural account of classification as inherently situated, challenging what Bourdieu (1991) calls the "symbolic violence" of imposing classificatory schemes that appear natural while serving particular interests. Hjørland's insight into the situated and value-laden nature of classification, articulated most programmatically in his 2002 synthesis of domain analysis, remains foundational for information science. His work foregrounds the epistemic commitments of communities and contests the presumed neutrality of universal classificatory systems—a critique that resonates strongly with the sociotechnical analyses developed in infrastructure studies, particularly by Susan Leigh Star and Geoffrey C. Bowker. Though arising independently in adjacent disciplinary contexts, both traditions emphasize the often invisible labor of classification and the infrastructural embedding of epistemic authority.

However, domain analysis assumes that communities possess sufficient internal coherence to generate stable classification norms, and that such norms can be expressed through classificatory logic. This assumption increasingly fails to hold in contemporary hybrid knowledge spaces, which are often marked by temporal friction—the emerging misalignment between infrastructural components that operate according to incompatible time horizons or epistemic rhythms.

What Hjørland identified as "domains" now often operate as what Manuel DeLanda (2006) calls "assemblages": epistemically significant clusters—tools, actors, platforms, institutions, and workflows—that function together without internal unity. These assemblages exhibit what DeLanda terms "exteriority"—their components maintain their own properties and can be detached and plugged into other assemblages, creating constantly shifting configurations that resist the stable boundaries that domain analysis assumes. This assemblage logic directly challenges domain analysis's assumption that knowledge domains have relatively stable boundaries and coherent internal structures. Following Latour's (2005) actor-network theory, these arrangements involve both human and non-human actors that together constitute hybrid epistemic agents capable of coordinating action across difference without requiring consensus.

**Metamorphic Logic of Concepts**

Epistemic infrastructures undergo conceptual metamorphosis as they encounter new contexts and challenges. Ideas are not merely transmitted unchanged but transform in ways that exceed their original design intentions. This insight draws from Karen Barad's (2007) "agential realism," which foregrounds the performative and relational enactment of knowledge, and extends to how concepts emerge through intra-active material–discursive processes.

However, SEI's framework also foregrounds a contrasting metaphysical stance, drawn from Graham Harman's (2018) object-oriented ontology, in which concepts—like all objects—withdraw from total relational access and retain autonomy beyond any enactment. SEI's diagnostic approach thus holds open a constitutive tension: between Barad's view of concepts as effects of entangled relations, and Harman's view of concepts as withdrawn entities that exceed any contextual deployment.

By preserving this tension, SEI enables a metamorphic logic of conceptual development: breakdown is not only a sign of failure, but also a site of conceptual mutation, where theoretical tools adapt, recombine, or exceed prior configurations. Theoretical responsibility, in this view, involves crafting concepts as epistemic objects—entities capable of enduring and transforming—rather than relying on fixed analytical instruments to guarantee continuity.

**The SEI Framework: Four Components of Situated Epistemic Infrastructures**

Contemporary epistemic assemblages function through co-functionality across difference—they work because their components do not necessarily agree, yet their outputs can be stitched together via mediating infrastructures. Building on Paul N. Edwards et al.'s (2013) analysis of knowledge infrastructures and Jean-Christophe Plantin et al.'s (2018) work on platform-infrastructure hybrids, SEI identifies four primary mechanisms through which knowledge is organized under post-coherence conditions:

- Material coordination through technical systems such as databases, repositories, and algorithms that determine how knowledge is stored, searched, and accessed. These affordances shape what Susan Leigh Star and Karen Ruhleder (1996) identify as infrastructure's reach and scope across temporal and spatial scales, creating dependencies that often remain invisible until breakdown occurs.
- Institutional embedding in organizational routines, funding models, and incentive structures that shape who produces knowledge, under what conditions, and with what forms of recognition. Following Paul J. DiMaggio and Walter W. Powell's (1983) analysis of institutional isomorphism, these embeddings create pressures toward similarity across organizations while potentially constraining innovative responses to changing conditions.

- Symbolic mediation through signs—"peer-reviewed," "evidence-based," "impact factor"—that serve as symbolic compressions of complex evaluative processes. Drawing from George Lakoff and Mark Johnson's (1980) work on conceptual metaphors and Ricoeur's (1984) analysis of narrative coherence, these symbols condition what counts as legitimate knowledge while often obscuring the complexity they represent.
- Governance and normativity through implicit norms about rigor, openness, transparency, and efficiency that are encoded in infrastructural design. Following Michel Foucault's (1991) analysis of "governmentality," these norms shape what is possible, permissible, incentivized, or discouraged, operating through what appears as technical neutrality while embedding particular assumptions about value and efficiency.

Situated Epistemic Infrastructures provides a framework for analyzing how knowledge is mediated, legitimated, and transformed through infrastructural arrangements under conditions of epistemic instability. Building on Star's (1999) foundational work on infrastructure studies, Mol's (2002) insights about ontological multiplicity, and the platform studies tradition (Gillespie, 2014; Plantin et al., 2018) the framework comprises four interlocking components that function as diagnostic lenses rather than rigid categories: infrastructure typology, power signatures, symbolic compression, and breakdown dynamics.

**Infrastructure Typology: Six Modalities of Epistemic Mediation**

Knowledge infrastructures operate through multiple, overlapping, and heterogeneous arrangements—an insight broadly developed in Star and Ruhleder's (1996) account of infrastructural complexity. Building on Edwards et al.'s (2013) analysis of knowledge infrastructures and extending it to address post-coherence conditions, SEI distinguishes six primary types based on how they mediate epistemic activity:

- Ritual infrastructures validate knowledge through what Collins (2004) analyzes as "interaction ritual chains"—ceremonial performances such as peer review, citation practices, and academic conferences that create emotional solidarity around community standards. They operate through repeated evaluative procedures that confer legitimacy by invoking established norms, requiring what Bourdieu (1986) terms "cultural capital" to navigate successfully. Their authority lies in symbolic enactment rather than technical verification, creating what Erving Goffman (1983) calls "interaction order" through carefully orchestrated performances that maintain status hierarchies while appearing to facilitate open evaluation. Examples include peer review processes across academic journals, the ritual of academic conference presentations, and citation practices that confer legitimacy through scholarly lineage.
- Institutional infrastructures formalize knowledge flows through organizational authority via journals, universities, funding agencies, and disciplinary associations. Drawing from Weber's (1978) analysis of legal-rational authority and DiMaggio and Powell's (1983) work on institutional isomorphism, they operate through rule-based governance that appears neutral while embedding assumptions about merit and efficiency. They control access to what Karl Maton (2003) calls "epistemic capital" through systematic procedures that create what Richard W. Scott (2014) identifies as "institutional logics"—coherent sets of material practices and symbolic constructions that provide organizing principles for organizational life.

- Communicative infrastructures include social media platforms, mailing lists, collaborative forums, and preprint servers that operate through iterative publicity—a mode of authority that builds through repeated cycles of sharing, commenting, and resharing, where legitimacy emerges from circulation patterns and network effects rather than credentialed authority. Following Jurgen Habermas's (1991) analysis of the public sphere and Tarde's (1903) insights about imitation and suggestion, they gain legitimacy through user engagement while potentially subordinating epistemic goals to what Nick Srnicek (2017) calls "platform capitalism"—the extraction of value from user interactions through attention capture and data collection.
- Algorithmic infrastructures such as search engines, recommender systems, citation metrics, and AI models shape knowledge circulation through automated logics. Their influence stems from computational scale and sophistication, generating epistemic dependence through mechanisms that remain largely opaque—a condition often described as "algorithmic authority" (Pasquale, 2015; Shirky, 2009; Simon, 2010). They embody what Langdon Winner (1983) identifies as "technological somnambulism"—the tendency to adopt new technologies without considering their broader implications for social relations and power distributions.
- Archival infrastructures stabilize knowledge via memory systems—repositories, libraries, metadata schemas—that structure how cultural memory persists across generations. Jan Assmann (1995, p. 126) defines cultural memory as "a collective concept for all knowledge that directs behavior and experience in the interactive framework of a society and one that obtains through generations in repeated societal practice and initiation." These infrastructures shape what is preserved and, just as critically, how it is practiced and reactivated over time. This stabilizing function is, however, also selective. As Derrida (1996) argues in his analysis of archival power, the archive exerts control over what is remembered and forgotten by means of curatorial logics that embed assumptions about value, authority, and relevance. The preservation of knowledge thus becomes inseparable from acts of exclusion.
- Experimental infrastructures represent ad hoc or emergent arrangements created to solve novel problems, including hackathons, crisis dashboards, and citizen science platforms. Their authority is context-dependent and provisional, derived from what can be termed crisis functionality—authority that emerges from proven effectiveness in addressing urgent problems when established infrastructures cannot respond quickly enough. They often emerge during breakdown moments and instantiate what Mol (2002) calls "ontological politics" by enacting new problem definitions and solution possibilities through experimental practices.

Each infrastructure type exhibits what Star (1999) calls infrastructural dimensions—such as embeddedness, transparency, scope, learned familiarity, and visibility upon breakdown—which normally remain invisible during routine operation but surface in moments of disruption. Hybrid combinations are now the norm, requiring analysis that can address co-functionality across difference—the capacity to coordinate diverse infrastructural logics without resolving them into unified approaches.

**Power Signatures: Patterns of Legitimation with Theoretical Vulnerability**

Each infrastructure type exhibits a distinctive power signature—patterns in how epistemic authority and recognition are distributed across networks of actors. Building on Bennett and Segerberg's (2013) analysis of political movements, we can identify how different knowledge infrastructures establish patterned ways in which authority is claimed and enforced through different modalities of power (Foucault, 1991). Extending Bourdieu's (1991) work on symbolic power to address contemporary technological mediation, SEI identifies several recurrent signatures that operate across modalities:

- Theoretical vulnerability. Genuine theoretical contribution requires a willingness to develop concepts that exceed one's capacity to control their interpretation or application. This concept of theoretical vulnerability represents a distinctive contribution to science and technology studies, suggesting that the most generative theoretical frameworks are those that can accommodate interpretations and applications exceeding their developers' intentions or control. The more generative a theoretical framework becomes, the more it will be taken up in ways that diverge from original intentions, potentially in directions that conflict with the developer's own theoretical or political commitments. SEI embraces this vulnerability as essential: the framework succeeds precisely when it becomes more than intended while remaining recognizably continuous with original insights. The concepts become conceptually autonomous—capable of generating insights and applications that exceed what their original developer could anticipate or control.
- Ceremonialism operates through ritual performance and cultural capital requirements, exemplified in journal peer review, academic rankings, and dissertation defenses. In this instance, authority derives from successful navigation of what Collins (1979) calls "credentialing" procedures that convert class advantages into apparently meritocratic achievements while requiring specialized knowledge and performative competence that effectively exclude those lacking appropriate cultural capital.
- Bureaucratic rationality exercises authority through formalized rules and procedures such as grant application protocols and university governance systems, embodying what Weber (1978) identifies as "legal-rational authority." Legitimacy comes from systematic application of formal procedures that appear neutral while embedding institutional assumptions about efficiency, accountability, and organizational effectiveness that may conflict with epistemic goals (Scott, 2014).
- Iterative publicity builds credibility through repeated public visibility and engagement, including social media circulation, comment threads, and media citations. Following Tarde's (1903) analysis of social influence and extending it to digital contexts, this modality enacts authority through circulation patterns and network effects rather than institutional credentials, enabling democratic participation while potentially subordinating epistemic goals to attention economies that often reward viral engagement over careful analysis.
- Algorithmic opacity lends authority to computational systems whose scale and sophistication foster dependence through technical mystification—exemplified by Google Scholar rankings and citation metrics, which are widely used despite operating through undocumented and easily gamed algorithms. As Pasquale (2015) notes, power in such systems operates through "black box" effects and the capacity to process information volumes that exceed human cognitive capacity, embedding assumptions about relevance and quality that remain hidden from users.
- Canonical memory exerts infrastructural power by shaping what is archived, indexed, and remembered—through mechanisms such as metadata standards, citation databases, and repository inclusion. Its authority stems from the ability to preserve knowledge across time while selectively curating what counts as worthy of remembrance. As Bowker (2005) notes, these "memory practices" embed cultural assumptions about relevance, systematically marginalizing alternative knowledge traditions through their omissions.
- Crisis functionality enacts a distinct form of authority grounded in exceptional performance during moments of infrastructural failure—such as COVID-19 dashboards and rapid response collaborations. Legitimacy arises from demonstrated utility under emergency conditions, where traditional validation procedures break down. This breakdown-generated authority challenges established hierarchies by demonstrating situational effectiveness and adaptive response rather than relying on formal credentials.

These power signatures exhibit a form of strategic reversibility—they may complement or conflict depending on how they are assembled in specific contexts. Most epistemic arrangements involve layered or contested signatures, where tensions between competing legitimation logics produce authority friction: a dynamic tension that can drive transformation or precipitate breakdown. This is evident, for example, in conflicts between COVID-19 dashboard transparency and the secrecy of traditional epidemiological modeling, or between arXiv's rapid circulation and the deliberative pace of journal peer review.

**Symbolic Compression: Coordinating Complexity**

Symbolic compression represents one of SEI's most distinctive theoretical contributions. Drawing on Lakoff and Johnson's (1980) theory of conceptual metaphors and Ricoeur's (1984) account of narrative coherence, the framework extends these insights to explain how complex epistemic processes are condensed into portable markers of legitimacy. These compressions function as epistemic currency—condensed signals that facilitate coordination across heterogeneous systems by encoding trust, quality, and authority into familiar terms. In contexts of information overload, they enable rapid decision-making while concealing the infrastructural and interpretive work on which their legitimacy depends.

However, following Bourdieu's (1991) analysis of symbolic power, they can obscure complexity, stabilize bias, or become sites of contestation when their compressed meanings become overloaded or strategically manipulated. SEI identifies three key mechanisms through which symbolic compression operates:

- Legitimation shortcuts enable gatekeeping by reducing complex evaluative processes to binary indicators that function as quality proxies. "Indexed in Scopus" serves as a proxy for journal quality, while "peer-reviewed" suggests comprehensive expert evaluation regardless of actual review procedures. Drawing from Susan Leigh Star and James R. Griesemer's (1989) concept of "boundary objects," these shortcuts enable efficient decision-making across different institutional contexts while potentially masking important variations in quality and process that become invisible through compression.
- Semantic overload occurs when compressions acquire multiple incompatible meanings across contexts, creating polysemic drift. "Open science" means different things to funders (transparency), activists (democratization), and institutions (efficiency), creating coordination problems when these different meanings conflict in practice. Similarly, "evidence-based" operates differently in medical, educational, and policy contexts with incompatible operational definitions, leading to ontological multiplicity—where the same term refers to different objects in different contexts.
- Symbolic drift occurs when compressions lose connection to their original practical commitments yet retain legitimating force, opening them to strategic appropriation (Hutchins, 1995). Commercial publishers adopt "open access" terminology while maintaining subscription models, and predatory journals claim "peer review" while providing minimal evaluation, demonstrating how symbolic authority can persist even when underlying practices are corrupted or abandoned.

Understanding symbolic compression is crucial because, while these processes enable functional coordination, they also create vulnerabilities when compressed meanings become semantically saturated—overloaded with conflicting interpretations that erode their coordinating capacity. The concept of peer review offers a clear example: originally denoting a complex editorial process involving multiple stages and actors, it now operates as a compressed signifier of legitimacy—facilitating rapid quality assessment while becoming vulnerable to symbolic drift when invoked by actors who retain the term but abandon substantive evaluative practice.

**Breakdown Dynamics: Diagnosing Infrastructural Failure**

Building on Star's (1999) insight that infrastructure often becomes visible through failure, and extending this through Charles Perrow's (1984) theory of "normal accidents" in complex systems, SEI treats breakdown as a diagnostic opportunity—exposing latent assumptions, invisible dependencies, and structural tensions through infrastructural inversion. Breakdown can occur along four dimensions that often interact through cascading effects where failures in one component generate stress in others:

- Technical breakdown involves material system failures that prevent normal operation, such as database outages, algorithm malfunctions, or platform crashes. These failures reveal technological dependencies while creating opportunities for alternative arrangements that might otherwise remain unthinkable, demonstrating what Winner (1980) calls the "political" dimensions of seemingly neutral technical systems.
- Symbolic breakdown is marked by erosion of trust, participant withdrawal, or contestation of authority—undermining infrastructural legitimacy through epistemic defection. When research communities abandon platforms perceived as biased or extractive, it signals a misalignment between user needs and platform logics, creating conditions for alternative arrangements to emerge through collective action.
- Social breakdown involves erosion of trust, participation withdrawal, or authority contestation that undermines infrastructural legitimacy through a process that may be termed epistemic defection. When research communities abandon platforms perceived as biased or extractive, it signals fundamental misalignment between user needs and platform logics, creating opportunities for alternative arrangements to emerge through collective action.
- Temporal breakdown occurs when emergent misalignments arise between infrastructural speeds and epistemic demands—such as when peer review timelines proved inadequate for pandemic response. These failures expose incompatible temporal assumptions embedded within different infrastructural logics, generating temporal friction between systems calibrated to divergent rhythms.

**Framework Integration: Diagnosing Rather Than Categorizing**

The four components of SEI (Table 1) offer a multi-dimensional model for analyzing epistemic infrastructures that operates through a diagnostic rather than classificatory logic—emphasizing situated enactments over static categories, in a manner consistent with the concept of ontological multiplicity:

**Table 1**: Four Components of SEI

| Component | Focus | Function |
| --- | --- | --- |
| Infrastructure Typology | Modalities of mediation | Identifies dominant forms and hybrid arrangements |
| Power Signatures | Patterns of legitimation | Reveals how authority operates and conflicts |
| Symbolic Compression | Coordination mechanisms | Explains meaning-making and strategic appropriation |
| Breakdown Dynamics | Failure and adaptation | Exposes assumptions, enables reflexive transformation |

Rather than pursuing exhaustive classification in the tradition of universal taxonomies, SEI advances a situated mode of analysis that poses contextually sensitive questions: Which infrastructural types are active in a given setting? Which power signatures dominate? Which symbolic compressions shape perception and decision-making? Where are breakdowns emerging or already visible?

This diagnostic flexibility enables SEI to analyze both local arrangements (institutional repositories) and macro-scale systems (global knowledge platforms) without collapsing their differences into unified models. It also enables comparative analysis that can address questions such as: How does knowledge organization differ when structured by ritual versus algorithmic infrastructures? What happens when symbolic compression fails in communicative systems? How do power signature conflicts create transformation opportunities?

The framework's strength lies in offering systematic tools for asking better questions about epistemic arrangements under conditions of permanent instability, rather than in providing predetermined answers. Following Dewey's (1938) pragmatist insight that tools function as instruments for inquiry rather than representations of reality, SEI provides analytical instruments that enable effective action under conditions of uncertainty and change.

## Large Language Models as Epistemic Crisis and Infrastructural Revelation

The emergence of large language models as epistemic infrastructures constitutes a computational epistemological rupture—an event in which knowledge production, validation, and circulation are being exposed, contested, and rapidly reconfigured through algorithmic mediation. Rather than treating LLMs as anomalous tools, SEI interprets them as accelerants and mirrors of broader epistemic transformation—forcing visibility on infrastructural assumptions long treated as implicit and demonstrating the diagnostic power of breakdown in the post-coherence condition.

Drawing once again from Star's (1999) insight that infrastructure becomes visible upon breakdown, this section applies the complete SEI framework to the LLM phenomenon (2022–present), demonstrating how the four components operate interactively under conditions of simulated epistemic coherence—where algorithmic systems generate seemingly authoritative responses by statistically recombining existing knowledge claims without the community-based validation processes that SEI identifies as fundamental to traditional epistemic legitimacy.

This analysis builds on Chen's (2025) empirical study of AI tools in education, which demonstrated how platforms like lesson plan generators and automated feedback systems, while allowing for sophisticated engagement, primarily market themselves for efficiency rather than epistemic enhancement. Chen's findings that such tools "inadequately support the skilled epistemic actions of teachers" and "potentially cultivate problematic long-term habits" exemplify the broader infrastructural tensions that SEI aims to diagnose across domains. Chen's three-dimensional framework—skilled epistemic actions, epistemic sensitivity, and habit-building—provides a complementary operationalization of infrastructural analysis that demonstrates how efficiency-focused design can systematically undermine epistemic agency even when preserving space for human judgment. Building on both Star's infrastructural insights and Chen's educational analysis, this examination proceeds through four systematic steps that reveal how computational infrastructural assemblages reconfigure knowledge work in real time: infrastructural convergence, power signature disruption, symbolic compression crisis, and reflexive adaptation.

**Infrastructural Convergence Under Computational Mediation**

The integration of LLMs into knowledge work has created what might reasonably be termed a *polytemporal computational zone*—a condition where human inquiry, algorithmic processing, and institutional validation must co-exist across radically misaligned temporal and epistemic rhythms. Traditional scholarly infrastructures operate through deliberative citation practices and peer review cycles, while LLMs generate responses instantaneously from training data with fixed temporal boundaries. This creates temporal friction between systems grounded in fundamentally different assumptions about evidence, rhythm, and revision (Bommasani et al., 2021; Brown et al., 2020).

While Star and Ruhleder (1996) emphasize the relational and often invisible nature of infrastructural breakdown, SEI extends their analysis to foreground the temporal misalignments that emerge when such systems interface. These frictions do not lead to exclusion; instead, they produce infrastructural convergence—a condition of forced cohabitation in which epistemically and temporally divergent systems must operate within shared workflows. This convergence is not seamless integration, but a tense alignment that reconfigures knowledge production under conditions of ambient epistemic instability.

This temporal desynchronization has triggered infrastructural convergence: disparate systems such as search engines, citation managers, academic databases, and conversational AI have become functionally coupled despite originating from incompatible epistemic logics. Following DeLanda's (2006) analysis of assemblages, these convergences produce novel computational-epistemic hybrids with emergent properties that exceed the sum of their components:

- Search engines, when combined with large language models, create emergent algorithmic-conversational assemblages that enable unprecedented query sophistication while undermining traditional mechanisms of source attribution. Platforms like Microsoft Copilot and Google Gemini integrate search retrieval with generative responses, producing compressed citation loops that turn complex bibliographic practices into conversational outputs, obscuring the infrastructural labor involved.
- Academic databases, when combined with AI assistants, integrate archival, algorithmic, and communicative infrastructures, as research practices are mediated by systems that process vast literature corpora but cannot reliably distinguish credible from unreliable sources (Bender et al., 2021). The use of AI writing assistants in academic work creates hybrid arrangements that combine what Weber (1978) calls "bureaucratic authority" with algorithmic plausibility, illustrating how distinct power signatures become entangled under computational conditions.

- Citation tools, when combined with generative AI, modify ritual infrastructures by automating traditionally ceremonial practices—such as bibliography generation and literature summarization—through systems that lack access to the full epistemic context of the cited works (Thorp, 2023). Research workflows increasingly involve AI-generated literature reviews that compress months of reading into minutes, producing simulated scholarship: the outward form of academic practice persists, but its epistemic labor is displaced by computational shortcuts.

This convergence compresses distinct epistemic temporalities into a polytemporal computational zone, where infrastructures built for human-paced deliberation must now accommodate machine-speed processing. While this temporal layering disrupts the pacing of scholarly judgment, it also signals a deeper transformation: the emergence of a *computational epistemic zone*, in which algorithmic infrastructures mediate, simulate, and restructure what counts as knowledge. The shift extends beyond acceleration; it signals the formation of hybrid epistemic regimes in which divergent infrastructural logics co-function under conditions of ambient uncertainty and simulated coherence. This process—termed algorithmic co-functionality—marks a transition from integration to entanglement: a mode of infrastructural interdependence that reconfigures both the rhythm and authority of knowledge production.

**Power Signature Disruption: Authority Under Computational Pressure**

LLMs make visible tensions between previously distinct power signatures, illustrating how Foucault's notion of "strategic reversibility" can be applied to the technological reconfiguration of power (Foucault, 1991). AI-generated content can appear authoritative when its confident form and institutional uptake obscure the underlying epistemic uncertainties of its production. This dynamic reverses the usual sequence in which credibility follows validation, unsettling conventional assumptions about how authority is conferred.

Ritual versus algorithmic conflicts emerge when peer-reviewed knowledge competes with AI-generated responses that synthesize information from sources that are undisclosed to users while presenting conclusions with apparent confidence. LLM outputs gain credibility through simulated coherence—the appearance of scholarly reasoning—while bypassing the ceremonial procedures that traditionally confer epistemic legitimacy (Ji et al., 2023). This generates authority conflicts that existing gatekeeping systems are structurally unequipped to resolve—a phenomenon of computationally induced epistemic bypass.

Institutional versus communicative tensions appear when university policies and professional standards lag behind the rapid adoption of AI in research and writing practices. Infrastructures grounded in bureaucratic rationality prove inadequate for governing systems that blend human and machine authorship, creating legitimacy gaps that erode traditional notions of scholarly integrity and expose the temporal assumptions embedded in human-centered authority frameworks.

Transparency versus opacity conflicts arise when LLMs used for research synthesis operate through proprietary algorithms and undisclosed training data, while claiming to provide comprehensive knowledge summaries. The epistemic tensions introduced by AI—where scholarly norms of traceability collide with algorithmic black-boxing—demand institutional responses grounded in what Sheila Jasanoff (2003) calls “technologies of humility,” which emphasize reflexivity, pluralism, and deliberation in the governance of expert knowledge.

These tensions signal a shift in authority and a redefinition of legitimacy. As traditional gatekeeping falters, credibility becomes entangled with computational systems that simulate coherence while bypassing established modes of validation. LLMs do not replace epistemic infrastructures—they fuse with them, forming hybrid regimes in which institutional, ritual, and algorithmic forms of authority uneasily co-exist under computational pressure.

**Symbolic Compression Under Computational Stress**

The integration of LLMs has intensified reliance on symbolic compression while simultaneously destabilizing these compression mechanisms through computational semantic drift. Terms like “AI-assisted,” “evidence-based,” and “comprehensive review” become high-velocity tokens that assert scholarly legitimacy even when underlying processes involve opaque algorithmic synthesis rather than traditional research methods, illustrating how computational mediation can accelerate and distort symbolic markers of legitimacy, enabling rhetorical authority without corresponding epistemic grounding.

AI-assisted research-compression stress arises when the term “AI assistance” obscures key distinctions—such as those between formatting and content generation, or between querying peer-reviewed versus general web-trained models (Cotton et al., 2023). This compression functions as a symbolic marker of technological sophistication, recalling Bourdieu’s analysis of how symbolic capital can obscure structural differences—particularly when assistance spans from spell-checking to literature synthesis without clear epistemic boundaries.

Comprehensive-analysis computational compression exemplifies how LLMs enable what appears to be exhaustive literature review through processing capacities that exceed human cognition, while obscuring the temporal limits of training data and the selection biases embedded in corpus construction. This compression condenses complex curatorial and analytical work into seemingly authoritative summaries that support rapid knowledge claims, while deferring questions of comprehensiveness—problems that arise when recent developments or marginalized perspectives are systematically excluded.

Evidence-based algorithmic appropriation occurs when LLM outputs claim empirical grounding while synthesizing information from sources that users cannot verify or access. Commercial AI systems adopt scholarly terminology to legitimize responses but often fail to support evidence-based claims—particularly in contexts requiring abductive reasoning (Dougrez-Lewis et al., 2023)—demonstrating how computational processing enables strategic appropriation of epistemic vocabularies while eroding the verification practices these terms originally signified.

These compression dynamics reveal how computational mediation enables coordination across vast information scales while introducing vulnerabilities when algorithmic synthesis diverges from traditional scholarly practices. As meanings fragment, terms like “hallucination” (Ji et al., 2023), “prompt engineering,” (Reynolds & McDonell, 2021)l and emerging ethical vocabularies around AI safety and “alignment” (Stahl & Eke, 2024) enter academic discourse to address challenges that existing conceptual frameworks cannot adequately capture. This illustrates the ongoing semantic labor required to sustain meaningful connections between computational capabilities and epistemic commitments.

**Breakdown Dynamics and Computational Adaptation**
SEI treats the LLM phenomenon as a reflexive computational opening—a moment when epistemic systems adapt to algorithmic mediation while creating precedents for hybrid human–AI arrangements that demonstrate possibilities for computational breakdown learning. Drawing from Chris Argyris and Donald A. Schön's (1978) work on organizational learning and extending it to address algorithmic contexts, breakdown dynamics operate across all four dimensions while catalyzing innovative responses:

- Technical breakdown manifests when LLMs generate plausible but inaccurate citations, produce outdated information due to training cutoffs, or synthesize contradictory sources without acknowledgment (Alkaissi & McFarlane, 2023; Ji et al., 2023). These failures reveal the infrastructural opacity of computational systems while catalyzing development of verification tools and hybrid workflows that can leverage AI capabilities while maintaining scholarly standards, demonstrating how breakdown drives computational infrastructural innovation when coupled with reflexive institutional response.
- Symbolic breakdown occurs through confusion over what constitutes "original" work when AI assists in writing and analysis, undermining traditional notions of authorship and intellectual contribution (Nature Editorial, 2023; Thorp, 2023). Terms like "scholarly integrity" lose unifying power as computational assistance makes traditional attribution practices inadequate, revealing the temporal and technological assumptions embedded in established ethical vocabularies while forcing reconsideration of human agency in knowledge production.
- Social breakdown becomes visible in withdrawal from AI-mediated platforms perceived as unreliable or extractive, and in conflicts over institutional policies governing AI use in academic work (Kasneci et al., 2023). Traditional authority relations shift as computational systems gain credibility through the appearance of expertise, even as human actors struggle to verify algorithmic claims. This dynamic gives rise to an epistemic hybridization in which authority emerges from human–AI collaboration rather than individual expertise.
- Temporal breakdown emerges from the misalignment between instantaneous AI responses and the deliberative pace of scholarly practice, producing cascading failures as researchers, institutions, and publishers attempt to reconcile systems built on fundamentally different temporal assumptions about knowledge validation (Brown et al., 2020). These disruptions reveal the temporal dependencies that bind computational and human infrastructures—dependencies that become visible precisely when algorithmic tempo collides with the rhythms of academic deliberation.

Reflexive adaptation responses include multiple innovations demonstrating computational infrastructural learning capacity:

- Institutional guidelines for AI use that distinguish between acceptable assistance and inappropriate substitution, maintaining scholarly values while enabling computational enhancement through ethical flexibility (Nature Editorial, 2023).
- Hybrid verification systems that combine AI synthesis capabilities with human oversight and source checking, enabling scalable analysis while preserving epistemic reliability through computational–human interoperability (Dwivedi et al., 2023).
- Transparency experiments including AI disclosure requirements and prompt sharing that provide alternatives to traditional citation practices through algorithmic accountability (Lund et al., 2023).

- Student and faculty repositioning where computational literacy becomes essential for scholarly practice, transforming relationships between human expertise and algorithmic capability through epistemic co-evolution (Kasneci et al., 2023).

These adaptations highlight that computational epistemic legitimacy emerges from ongoing negotiation—developing through human–AI co-functionality rather than fixed technological solutions. The rise of large language models makes infrastructural mediation visible to both experts and everyday knowledge workers, reconfiguring relationships between human cognition, computational processing, and institutional validation in ways likely to persist as AI capabilities continue to advance.

**What LLMs Reveal About Computational Epistemic Infrastructure**

Rather than simply a case study in technological adoption, the LLM phenomenon should be understood as a revelatory event that exposes the computational mediation of contemporary epistemic life while demonstrating possibilities for hybrid human–AI arrangements that extend Star's insights about infrastructural visibility (Star, 1999). The phenomenon reveals several key insights about how epistemic infrastructures operate under computational conditions:

- Legitimacy emerges from algorithmic–human coordination, not computational autonomy. Authority develops through successful integration of AI capabilities with human oversight rather than from algorithmic sophistication alone, showing how computational epistemic legitimacy depends on hybrid networks of association between human expertise, institutional validation, and algorithmic processing rather than technological replacement of human judgment.
- Breakdowns reveal computational epistemic limits. AI failures create opportunities for developing better human–machine collaboration by exposing the boundaries of algorithmic reliability (Ji et al., 2023). demonstrating the importance of building adaptive capacity into computational workflows rather than optimizing for seamless automation.
- Power signatures hybridize through computational mediation. LLMs show how algorithmic and human authority modalities can complement each other while revealing possibilities for new forms of epistemic legitimacy when computational opacity is managed through transparency coordination that preserves both efficiency and accountability.
- Symbolic compressions face computational amplification. AI systems accelerate both the utility and the fragility of epistemic shortcuts, highlighting the importance of maintaining computational semantic stewardship—ongoing work to preserve meaningful connections between algorithmic capabilities and scholarly practices while enabling rapid knowledge processing.
- Epistemic authority requires infrastructural literacy. Effective use of LLMs for knowledge work demands understanding of their training biases, temporal limitations, and synthesis mechanisms, suggesting new forms of computational epistemic competence that integrate technical knowledge with scholarly judgment.

SEI's analytical power lies in its ability to frame these dynamics as enduring characteristics rather than temporary technological disruptions, revealing defining features of computationally mediated epistemic life. The framework illuminates how algorithmic infrastructures actively reshape the conditions of knowledge production while demanding reflexive human engagement to preserve epistemic values. In doing so, SEI provides a foundation for examining how knowledge organization must adapt in an era of ubiquitous AI assistance.

## Implications and Conclusion: Toward Reflexive Epistemic Futures

Situated Epistemic Infrastructures represents more than theoretical innovation—it signals a fundamental reorientation in how knowledge organization approaches its mission under conditions of permanent instability. Drawing from Dewey's (1938) pragmatist insight that thinking is reconstruction rather than representation, and extending it through contemporary work on reflexive modernization (Beck et al., 1994), this concluding section examines SEI's implications across five dimensions: theoretical contribution, methodological innovation, applied relevance, future directions, and reflexive ethos.

### From Classification to Coordination: Theoretical Reorientation

SEI shifts the foundational concern of knowledge organization from classification to infrastructural coordination, building on the "infrastructural turn" in science and technology studies while extending it to address post-coherence conditions. Traditional models sought to represent domains, codify consensus, and stabilize knowledge through systematic categorization—an approach that Barad (2007) critiques as representationalist, insofar as it presumes that knowledge merely reflects a pre-given reality. While Barad theorizes *material–discursive practices* to emphasize the inseparability of matter and meaning, SEI adopts the related notion of *material–symbolic arrangements* to highlight how sociotechnical infrastructures—including classification systems and computational platforms—mediate epistemic activity. This orientation supports what SEI terms coordinative mediation: the ongoing work of enabling co-functionality across epistemic difference without requiring consensus or assimilation into unified frameworks.

This theoretical move introduces several core insights that reframe how we understand epistemic organization:

- Legitimacy is not produced by coherence but by convergence**.** Epistemic authority arises from the ability of infrastructures to coordinate across difference through what can be termed situated coordination, rather than by enforcing sameness within bounded communities. The COVID-19 preprint ecosystem gained legitimacy through effective coordination among researchers, policymakers, and publics—not through disciplinary consensus—demonstrating that authority can emerge from functional alignment rather than cultural agreement.
- Breakdown is normal. Rather than treating infrastructural failure as exceptional, SEI positions breakdown as what Perrow (1984) calls "normal accidents" in complex systems—a key moment of visibility, reflexivity, and transformation that reveals normally hidden dependencies and assumptions. Understanding how infrastructures fail provides crucial insights into how they operate and how they might be improved through breakdown learning.
- Situatedness is infrastructural. Domain specificity still matters, but following Haraway's (1988) insight into situated knowledges, it is better understood as one dimension within broader assemblages of technological, institutional, and symbolic relations. Communities operate within infrastructural conditions that both enable and constrain epistemic possibilities through what may be termed embedded mediation (Graffagnini et al., 1999).
- Power operates through mediation. Authority relations are embedded in infrastructural arrangements rather than simply reflecting community hierarchies or institutional positions, following what Foucault (1991) calls "governmental" rather than "sovereign" modalities of power. Understanding power signatures reveals how different legitimation logics compete and interact through authority dynamics that shape epistemic possibilities.

SEI reframes knowledge organization as an ongoing practice of epistemic cultivation—a dynamic negotiation of tensions among local stability, adaptability, and justice within a context of persistent infrastructural volatility.

**Embracing Infrastructural Reflexivity: Beyond Coherence**

The most far-reaching implication of SEI is epistemological: it repositions infrastructure as a central object of reflexive attention, drawing from Beck et al.'s (1994) work on "reflexive modernization" and extending it to epistemic contexts. To organize knowledge today is to design and maintain infrastructural systems that are simultaneously fragile and consequential, often invisible until they fail. SEI insists that epistemic practice includes infrastructural awareness—that knowledge work is shaped by argument, evidence, mediation, compression, and coordination through material–symbolic arrangements that require ongoing attention and care.

Rather than offering a blueprint for coherence, SEI cultivates comfort with productive tension between competing values and requirements that cannot be resolved through technical solutions alone:

- Between fast and slow knowledge production that requires temporal flexibility rather than uniform pacing.
- Between institutional control and platform dynamism that demands governance innovation capable of managing hybrid arrangements.
- Between technical order and symbolic ambiguity that necessitates semantic stewardship to maintain meaningful connections between symbols and practices.
- Between community traditions and global coordination that calls for situated universalism—approaches that can coordinate across difference without imposing uniformity.
- Between stability and adaptive transformation that requires dynamic equilibrium rather than static solutions.

The goal is to cultivate capacities for navigating disorder wisely—developing analytical tools, diagnostic methods, and design strategies suited to the post-coherence condition of contemporary epistemic life rather than pursuing the restoration of traditional epistemic order. This requires developing new forms of professional expertise that combine technical knowledge with critical analysis, democratic engagement, and adaptive learning through infrastructural literacy—the capacity to recognize, analyze, and intervene in the material–symbolic arrangements that shape epistemic possibilities.

SEI suggests that the future of knowledge organization lies in cultivating ongoing capacities for diagnosing, adapting, and improving infrastructural arrangements that can support human flourishing under conditions of permanent change. The framework provides theoretical foundations and practical tools for this ongoing work while recognizing that the most important task—applying these tools thoughtfully and adaptively to address real challenges in specific contexts—remains ahead.

Most fundamentally, SEI points toward possibilities for epistemic organization that can support both rigorous inquiry and democratic participation, both local autonomy and global coordination, both innovation and sustainability—through ongoing cultivation of reflexive, adaptive, and justice-oriented epistemic arrangements rather than predetermined solutions. Drawing from Dewey's (1938) insight that the path of inquiry is more important than final answers, the framework contributes to broader efforts to develop knowledge systems that can serve the challenges of the 21st century while preserving the values of critical inquiry and democratic participation that make knowledge work meaningful. In this way, SEI offers a new theoretical framework and a diagnostic orientation toward epistemic life under post-coherence conditions—an approach that can help communities develop the analytical capabilities and practical tools needed to consciously cultivate the infrastructural arrangements through which knowledge is produced, validated, and mobilized in service of human flourishing and planetary sustainability.